\documentclass[preprint2]{aastex}		

   \shorttitle{Three Emission-Line Galaxies at \mathbf{z \sim 2.4}}
   \shortauthors{Meyer et. al.}
   
\begin{document}

\title{Three Emission-Line Galaxies at $\mathbf{{\it z} \sim 2.4}$}

\author{Richard Meyer\altaffilmark{1}}
\affil{California Polytechnic University, San Luis Obispo, CA}
\email{rmeyer@calpoly.edu}

\author{David Thompson}
\affil{California Institute of Technology, Pasadena, CA}
\email{djt@irastro.caltech.edu}

\and 

\author{Filippo Mannucci}
\affil{C.A.I.S.M.I.---C.N.R., Florence, Italy}
\email{filippo@arcetri.astro.it}

\altaffiltext{1}{Summer Undergraduate Research Fellow, California
  Institute of Technology, Pasadena, CA}

\begin{abstract}
We present Keck near-infrared and WIYN optical photometry of a sample 
of galaxies detected by near-infrared narrowband imaging in the fields 
of quasar metal absorption line systems at $z\sim2.4$.  Wide separations 
(0.6-1.6\,$h^{-1}$ Mpc) from the quasars indicates that they are not
directly responsible for the absorption systems.  From the color excess
of the galaxies we derived line fluxes, star formation rates, and
equivalent widths.  The data are consistent with one source having an active
nucleus and two sources containing regions of star formation. The blue
(R-K) colors for the sources suggest relatively lesser dust content.
We discuss possible projects using current wide-field infrared
instruments, which can cover an order of magnitude greater area with 
modest allocations of telescope time.   
\end{abstract}


\section{Introduction}

A complete understanding of the physical processes involved in the
formation of galaxies requires a knowledge of the young galaxy populations
at high redshift.  These primeval galaxies are thought to exhibit
bursts of star formation possibly associated with active galactic nuclei 
(AGNs), detectable through the resulting strong emission lines.  Various
detection techniques have been explored over the years in search of
these star forming galaxies.  Each detection technique introduced has
both advantages and disadvantages, and therefore it is important to
explore many different techniques to obtain a view of the primeval
galaxy population as complete as possible.

The Lyman-alpha emission line has been the target of many surveys
aimed at the detection of high redshift galaxies (see reviews by 
\citet{p94,tmfhrb98}), but turned up only a small fraction of the 
expected number of strong emission line sources.  More recent deep 
surveys \citep{ch98,hu98,hu99,pwk98,k00} have detected some Ly$\alpha$ 
emitters, but at lower luminosity.  The Large-Area Lyman Alpha survey 
\citep{rhoads00} currently being conducted has turned up a number of 
Ly$\alpha$ emitters, indicating that these sources are detectable with 
an adequate volume coverage and sensitivity.

An alternative method, which does not rely on the presence of a
strong Ly$\alpha$ line, is the Lyman Break technique 
\citep{steidel92,steidel93,steidel95}.  This technique involves the
detection of both a blue UV continuum and the Lyman limit spectral 
drop at a rest-frame 912\,\AA.  This technique has proved very 
effective at identifying $3 \leq z \leq 3.5$ galaxies.  Spectroscopic 
analysis \citep{steidel96} has shown a near zero mean Ly$\alpha$
emission, which is consistent with the lack of strong Ly$\alpha$ 
emitters in the line-based surveys.  Highly reddened sources would 
not be selected by the Lyman Break technique. 

High redshift ($z \geq 1$) near-infrared emission-line surveys search for
rest-frame optical emission lines, such as H$\alpha$ $\lambda$6563, 
[O~III] $\lambda$5007, H$\beta$ $\lambda$4861, or [O~II] $\lambda$3727.  
The attenuating effects of dust are far less at longer wavelengths, so 
these emission lines can detectable in dustier galaxies.  Ground-based 
emission line surveys have primarily targeted fields containing objects 
with known redshifts ($z > 2$), typically quasars and radio galaxies 
\citep{b95,pd95,tmb96}, or damped Ly$\alpha$ systems and strong metal 
absorbers, \citep{tmm98,mtbw98}, but also blank fields \citep{mvco00}. 
In addition, \citet{mcc99} used the Hubble Space Telescope NICMOS
camera to conduct a parallel, slitless spectroscopic survey.  Targets 
are selected at specific redshifts which place the strong optical lines 
into the bandpasses of existing near-infrared narrowband filters. 
\citet{malkan95} and \citet{btmd98} spectroscopically confirmed 
two emission-line galaxies.

In this paper we present followup observations on the 18 candidate
emission-line objects from \citet[hereafter MTBW98]{mtbw98},
identified in a survey covering 227\,arcmin$^2$.  We obtained both
optical and near-infrared observations on the majority of their sources.  
Our goal was to investigate the morphology, color, and confirm the 
presence of the emission lines.  In \S\ref{obs} we describe the 
observations and reduction methods.  In \S\ref{results} we discuss the 
results, including possible reasons for why many of the objects were 
not confirmed. Finally, in \S\ref{sum} we summarize our results and 
discuss prospects for future surveys using a narrowband near-infrared 
imaging technique.  For simplicity and to ease comparisons with 
prior surveys, we assume a cosmology of $H_{0} = 100$\,$h$ 
km\,s$^{-1}$\,Mpc$^{-1}$, $\Omega_{\rm M} = 1$ and $\Omega_{\Lambda} = 0$.  
Assuming a Lambda cosmology, with $H_{0} = 70$\,km\,s$^{-1}$\,Mpc$^{-1}$, 
$\Omega_{M} = 0.3$ and $\Omega_{\Lambda} = 0.7$, would increase 
projected separations by a factor of two, or the derived star formation 
rates or areas surveyed by a factor of four.

\section{Observations \& Reductions} 
\label{obs}

\subsection{Optical Imaging}

Deep $R$ band images covering 12 of the 18 emission-line
candidates from MTBW98  were acquired with the CCD imager in the queue 
observing mode of the 3.5m WIYN telescope from August to November
1997. The imager uses a 2048$^{2}$ pixel CCD at 0\farcs195 per
21$\mu$m pixel, resulting in a 6\farcm66 square field of view. 
The data were obtained using a Harris R filter with a central
wavelength of 646nm and a full width at half maximum (FWHM) of 153nm.  

Standard CCD processing was used to bias-subtract, flatfield, mask the
hot pixels, align, and then stack the images. The seeing in the data were
good, with an average 0\farcs62 FWHM.     

The deep observations were not obtained under photometric conditions,
so calibration images were acquired for each field during photometric 
weather on subsequent nights. The photometric images were calibrated 
onto the Vega scale using the \citet{landolt92} standards.  The images 
reach limiting (5$\sigma$) magnitudes between 25\fm7 and 26\fm4.  
 
\subsection{Infrared Imaging}

Infrared images were taken in either the $J$ or $K$ band, and in the 
corresponding 1.237$\mu$m or 2.248$\mu$m narrowband (NB, 
$\delta\lambda / \lambda \sim 0.01$) filters for 15 of the 18 sources 
from MTBW98. Images were obtained on UT 1999 July 22 and 23 with the
Near-Infrared Camera \citep{nirc94} on the Keck I telescope.   NIRC uses 
a 256$^{2}$ InSb detector at 0\farcs15 pixel$^{-1}$ resulting in 
38\farcs4 square field of view.  The data consisted of 10-15 dithered 
images for each object, with 30s exposures for the broadband (BB) data 
and 180s for the narrowband data.  Thus, typical exposure times were 
5 minutes in the broadband filters and 30 minutes in the narrowband filters.  
The seeing ranged from 0\farcs3 to 0\farcs5 FWHM in $K$ and 0\farcs5 
to 0\farcs6 in $J$.  All observations were obtained under photometric 
conditions.

Using IRAF\footnote{IRAF is distributed by the National Optical
   Astronomy Observatories, which are operated by the Association of
   Universities for Research in Astronomy, Inc., under cooperative
   agreement with the National Science Foundation.}
routines, images were dark subtracted and flatfielded using a combination 
of domeflats and skyflats.   The bad pixels were masked using separate 
narrowband and broadband bad pixel masks because the longer exposure times 
for the narrowband images resulted in more saturated hot pixels.  Images 
were then aligned and stacked.  

The broadband data were calibrated onto the Vega scale using the 
\citet{pers98} standard stars. The narrowband zero-points were scaled to 
the broadband data to give the continuum objects in a given filter set a 
zero mean color for the night.  Photometry on the objects was extracted
in a 2$^{''}$ diameter aperture.  This relatively small aperture was
used to minimize contamination from nearby objects.  The photometric 
uncertainties quoted in Table~\ref{tbl-1} are determined by the IRAF/apphot 
package from a local measure of the sky noise around each source.  The 
uncertainties are not corrected for errors in determining the zero points, 
which we estimate at 3\% for the broadband data and 10\% for the narrowband 
data.  This has no effect on the relative colors of the sources.  

\section{Results \& Discussion}\label{results}
 
We constructed color magnitude diagrams for each field, plotting the 
(BB-NB) color against the narrowband magnitude.  Continuum objects appear 
at zero color because of the relative photometric calibration of the 
narrowband images.  An emission-line object will have a positive color 
since the total flux of the narrowband image will be due mostly to the
emission-line and not the continuum.

In addition, the broadband and narrowband image pairs were ``blinked'' 
in order to visually confirm the sources with excess narrowband flux as 
well as search for new emission line galaxies.  The images were displayed 
such that the continuum objects in the field had the same apparent 
brightness in both images, thus emission-line sources appear brighter 
in the narrowband image relative to the broadband.  Of the fifteen 
emission-line candidates from MTBW98 we confirmed strong emission in 
three sources, with (BB-NB) colors ranging from 0.75 to 2.22 magnitudes.  
No new emission-line sources were identified.

\begin{figure}[!t]
   \plotone{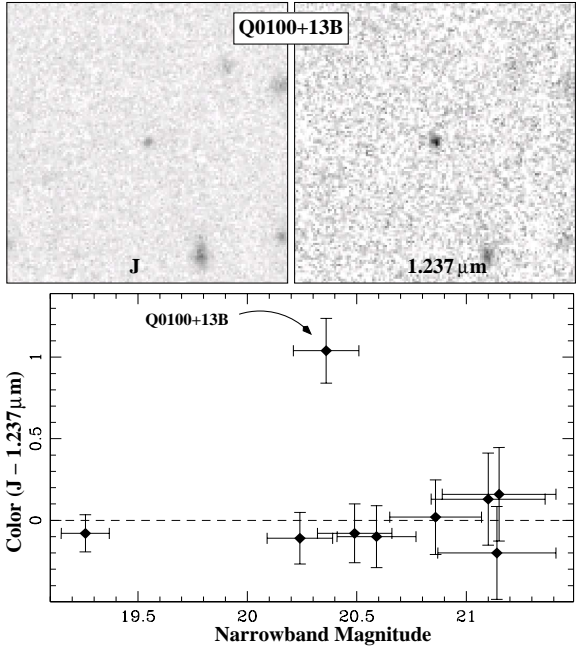}
   \caption{\small Top: $J$~band (left) and 1.237$\mu$m narrowband
      (right) NIRC images of Q0100+13B (centered in both images).  
      The J~band image has been scaled so that the continuum objects 
      appear the same in both images. Images are 20$^{''}$ square, with 
      north up and east to the left.  Bottom: color magnitude diagram of 
      the full ~60$^{''}$ field around Q0100+13B.
   \label{fig1}}
\end{figure}

\begin{figure}[!t]
   \plotone{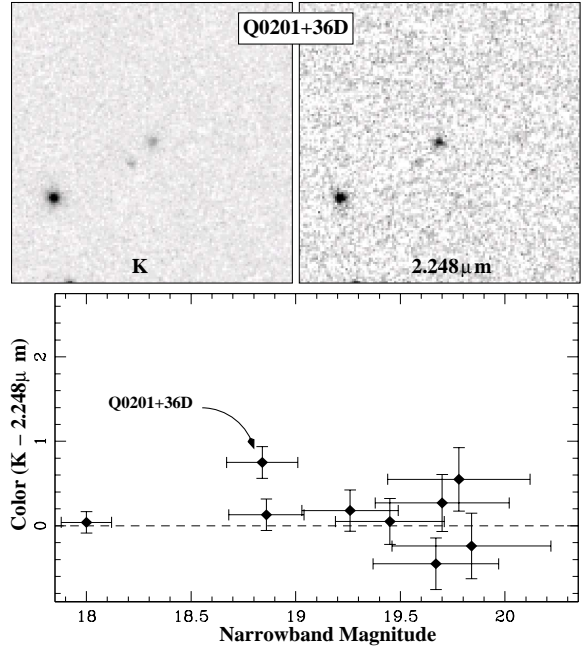}
   \caption{\small As in Fig.~\ref{fig1}, $K$ band and 2.248$\mu$m narrowband 
      images with color magnitude diagram of the field.  
   \label{fig2}}
\end{figure}

\begin{figure}[!ht]
   \plotone{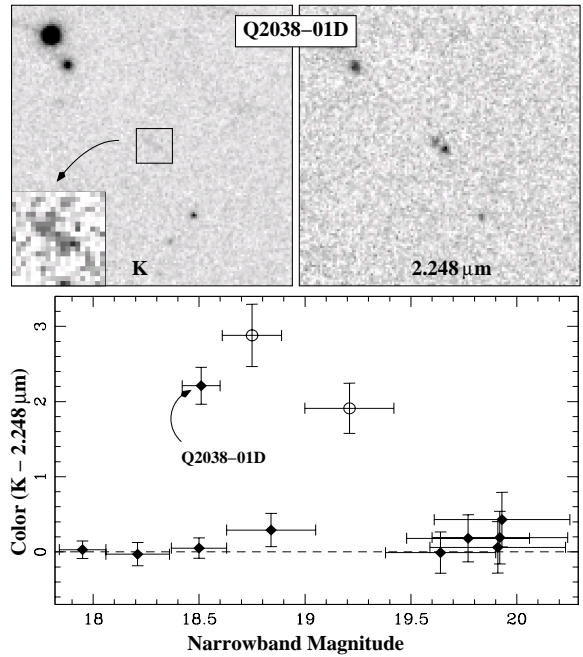}
   \caption{\small As in Fig.~\ref{fig1}, $K$ band and 2.248$\mu$m narrowband
      images for Q2038-01D, with a color magnitude diagram for the full
      field.  The open circles mark the photometry derived separately for 
      the two components.  The inset shows in more detail the 3$^{''}$ box 
      around the $K$-band image of Q2038-01D.  
   \label{fig3}}
\end{figure}

In the following sections we present notes on the individual sources. 
Table~\ref{tbl-1} lists the optical and near-infrared magnitudes and 
uncertainties for each source, as well as values derived from the color 
excess, such as the line equivalent widths, line fluxes, and star 
formation rates (SFR).  Figures~\ref{fig1}-\ref{fig3} show broadband 
and narrowband image pairs for each confirmed emission-line source and 
a color magnitude diagram.  The color magnitude diagrams were plotted 
from larger images ($\sim$60$^{''}$ square), and therefore contain data 
for more sources than are visible in the 20$^{''}$ square images.

\subsection{Q0100+13B}

The field of quasar Q0100+130 (PHL 957) was targeted by MTBW98 due to
the presence of a strong C\,{\sc iv} absorption line system at $z=2.308$
\citep{york91}, which places the redshifted [O~II]~$\lambda3727$ line at 
1.237$\mu$m.  The emission line candidate Q0100+13B shows a strong 
emission-line with a (J-1.237$\mu$m) color excess of 1.04$\pm$0.20 
magnitudes.  We derived a rest frame equivalent width of 60\,\AA\ and 
a line flux of $1.61 \times 10^{-16}$ ergs cm$^{-2}$ s$^{-1}$.  Assuming 
the emission-line is [O~II]~$\lambda3727$ at $z=2.31$, we used an [O~II] 
to H$\alpha$ line ratio in \citet{kennicutt92} with no correction for 
extinction to derive a star formation rate of 31\,$h^{-2}$ M$_{\odot}$
yr$^{-1}$.  If significant dust extinction is present, this star formation 
rate is a lower limit.  

Q0100+13B is separated from the quasar by 51\farcs2 at a position
angle (PA) of 35\fdg6.  The corresponding projected comoving separation 
from the quasar is 668\,$h^{-1}$~kpc (202\,$h^{-1}$~kpc proper separation).  
This is much larger than the size of an individual galaxy.  Since this 
source has the lowest projected separation of the three confirmed 
emission-line objects from their corresponding quasar's line-of-sight, 
none of these galaxies are likely to be responsible for the absorption-line 
systems targeted by MTBW98. 

Q0100+13B is marginally resolved with a 0\farcs6 FWHM in the
continuum subtracted image, however there are no good point sources
in the NIRC image to directly compare to and we used a seeing
average of 0\farcs5 FWHM for the night in the $J$ filter.
The moderate equivalent width and the slightly resolved 
morphology make Q0100+13B more consistent with the interpretation
that the line emission is powered star formation, but with the
current data we can not rule out an active nucleus.

\subsection{Q0201+36D}

The quasar Q0201+365 was targeted because of a strong C\,{\sc iv} 
absorption line system at z=2.424 \citep{sargent89,york91}, which 
places the redshifted H$\alpha$ at 2.248$\mu$m.  We confirm strong 
emission from the candidate Q0201+36D, with a (K-2.248$\mu$m) color 
excess of 0.75$\pm$0.19 magnitudes.  We derived a rest frame equivalent 
width of 67\,\AA\ and a line flux of $1.29 \times 10^{-16}$ ergs 
cm$^{-2}$ s$^{-1}$.  Assuming the line is H$\alpha$ at $z=2.42$, we 
used the luminosity to H$\alpha$ ratio from \citet{kennicutt92} to 
derive a SFR of 12\,$h^{-2}$ M$_{\odot}$ yr$^{-1}$.  Q0201+36D is 
separated from the quasar by 116\farcs3 at PA=141\fdg1. The corresponding 
projected comoving separation is 1.55\,$h^{-1}$ Mpc.  

The seeing in the Q0201+36D field was 0\farcs4 FWHM.  A profile fit to 
the continuum subtracted image failed to converge, so we resorted to a 
more subjective assessment of the source resolution.  A bright point 
source was extracted from elsewhere in the image, scaled down to a 
comparable flux as the emission line source, and then shifted and 
added back into the image next to the emission line source.  A visual 
comparison of the two shows that Q0201+36D is clearly resolved. 
If the emission is H$\alpha$ at a redshift 2.42, then the relatively 
low equivalent width and star formation rate, along with the diffuse 
nature of the emission line flux, suggest that this source is a star 
forming galaxy and not an AGN. 

\subsection{Q2038-01D}

As with the other two sources, the quasar Q2038-012 was targeted due to 
a strong C\,{\sc iv} absorption line at z=2.424 \citep{sargent89,york91}.
We confirm strong emission in candidate Q2038-01D in the 2.248$\mu$m
narrowband filter.  Our data show that Q2038-01D is resolved into two 
compact sources, both with emission lines, separated by 0\farcs8 
(10\,$h^{-1}$ kpc).   By masking out one source and measuring the 
photometry on the other, we measured the north-east (NE) and south-west 
components to have a color excess of 1.91$\pm$0.41 and 2.88$\pm$0.46 
magnitudes, respectively.  The derived rest-frame equivalent widths are 
quite high, over 10$^4$\AA\ for the SW component under the assumption 
that the line is H$\alpha$ at the targeted redshift.  Such a high 
equivalent width would be unusual even for an active nucleus.  There 
remains the possibility that this source is at a higher redshift, perhaps 
seen in the [O~III] or [O~II] line.  Spectroscopic data covering one or more 
additional lines could resolve this issue.  We note that the equivalent 
width becomes highly non-linear for a ($K - 2.248\mu$m) color in excess 
of two magnitudes, going to infinity at 2\fm96 where all of the broadband 
flux is from the emission line.  Q2038-01D is 120\farcs8 from the quasar 
at PA=126\fdg5.  The corresponding projected comoving separation is 
1.60\,$h^{-1}$ Mpc.  

The seeing in these data were especially good with an average of 0\farcs3
FWHM.   We measured the FWHM of each object by first masking out the other 
source.  Both components were formally resolved, with a 0\farcs45 FWHM, 
but the SW component appears more compact.  Faint, diffuse, elongated 
emission is visible on both sides of these sources in the narrowband image. 

Q2038-01D is most likely two galaxies in the process of merging, where 
the gravitational interaction is fueling perhaps both star formation 
and an AGN.  While the rest-frame equivalent width in the NE component 
is possible to produce with a pure starburst, the much larger equivalent 
width in the SW component almost surely indicates the presence of an 
active nucleus.  
 
\subsection{Source Colors}

All three confirmed emission-line objects had bluer (R-K) colors than 
the unconfirmed objects from MTBW98.  The (R-K) colors\footnote{We 
converted Q0100+13B's R-J color of 2.2 to R-K assuming a J-K of 1.} 
for the three confirmed sources range from 2.97 to 3.28 magnitudes.  
The mean color for a field sample at similar K-band magnitudes is 3.93 
\citep{thompson99}, indicating that these emission-line objects may not 
suffer from heavy dust obscuration, and it would be interesting to see 
if the rest-frame UV lines are visible with visual wavelength spectroscopy.  
While the original survey (MTBW98) should be sensitive to emission-line 
sources with more dust extinction than the UV techniques, in practice any 
objects with emission lines in the narrowband filter would be identified.  
This may indicate that very dusty sources are rare, but the number of 
objects found is still small.

\subsection{Unconfirmed Sources}\label{why}

Twelve of the fifteen candidates from MTBW98  did not show signs of an
emission-line in our data.  We propose two possibilities to explain this.
The most straightforward scenario is that the objects are not truly 
emission-line sources, but spurious detections in the narrowband
survey data.  The other scenario is that instrumental idiosyncrasies
could combine to cause effective shifts between the actual band-passes
of the narrowband filters.  

Each candidate object in MTBW98 was given a rank degree of significance, 
with 1 as the highest and 3 as the lowest.  All three of our confirmed 
sources had a rank of 1 or 2, while the unconfirmed candidates had ranks 
of 2 or 3.  Each of our confirmed objects has the highest signal to noise 
ratio (SNR) of all objects in the same field, ranging from 3.0 to 
5.6$\sigma$ in MTBW99, and the SNR was very low for many of the other 
sources.   So perhaps the emission-line candidates are simply 
spurious detections, selected from noise peaks in the narrowband
survey data.  We consider this the most likely explanation.

The other possibility for a lack of emission-line detection is due to
effective bandpass shifts between the narrowband filters used in the
survey and this followup.  The actual bandpass of the narrowband
filters, as seen by individual objects, is a function of both its
relative position in the field and any tilts of the filters in the
pupil planes of either instrument. These effects are difficult to
quantify a posteriori.  By using two separate filters with these
unknown variables, an emission line may be within the bandpass
of one filter, but not necessarily the other. Therefore we cannot 
conclusively rule out the presence of emission lines in our twelve 
unconfirmed sources.  
  
\section{Summary}
\label{sum}

We observed 15 candidate emission-line sources from MTBW98 and
confirmed strong emission in three.  These galaxies may be associated
with quasar metal absorption line systems at $z\sim2.4$, however, wide
separations (0.6-1.6\,$h^{-1}$ Mpc) from the quasars indicates that they 
are not directly responsible for the absorption systems.   While it is 
likely these galaxies are at the targeted redshifts, confirmation requires 
spectroscopic observations of one or more additional emission lines.
From the color excess of the galaxies we derived line fluxes, star 
formation rates, and equivalent widths.  Source resolution and line 
equivalent widths in two galaxies are consistent with line emission 
powered by star formation, while the third galaxy, which is actually 
double, is consistent with the presence of an active nucleus.  The 
blue (R-K) colors for the sources, as compared to field galaxies, suggest 
a relatively low dust content.  These galaxies may therefore be detectable 
in Ly$\alpha$ or other rest-frame UV emission lines.  

\subsection{Future Prospects}

A new generation of near-infrared instruments with larger fields 
of view are now available. Here we examine the feasibility of
conducting new narrowband infrared surveys.  

Several new, wide-field near-infrared instruments are currently working or
nearing completion on 4m class telescopes.  Here we specifically
consider Omega-2000 \citep{o2k} at the Calar Alto 3.5\,m telescope, 
with a 15.4 $\times$ 15.4 arcmin$^{2}$ field of view, WIRC-2k 
\citep{wilson02} at the Palomar 200-inch telescope with a 8.3 $\times$ 
8.3 arcmin$^{2}$ field of view, and FLAMINGOS \citep{flamingos} at the 
Kitt Peak 4\,m telescope with a 10 $\times$ 10 arcmin$^{2}$ field of view.  
MTBW98 reached line flux limits of $\sim2 \times 10^{-16}$ ergs 
cm$^{-2}$ s$^{-1}$, which we consider to be a conservative estimate for 
the new cameras, assuming 2 hour narrowband plus 30 minute broadband 
exposure times.  Under these assumptions, new narrowband surveys can cover 
an order of magnitude more area than MTBW98 in just 3-10 nights, a 
moderate-sized project, especially when split over more than one semester.  
Note that the large fields of view cover projected comoving fields of up to 
12\,$\times$\,12\,$h^{-2}$ Mpc$^{2}$ at z=2.4, much larger than the size 
of a cluster core.  Even surveys targeting known objects would then largely 
sample blank fields.  Removing the restriction of where to point, a 
purely blank field survey is free to target regions with low Galactic 
dust extinction and use the narrowband filters with maximum transmission 
or minimum background to increase the survey depth.  Additional survey 
volume can be covered at the expense of reducing the color contrast by 
using somewhat broader narrowband filters (e.g. $\delta\lambda / \lambda 
\sim 2-3$\%).  

For 8\,m-class telescopes, currently operating near-infrared instruments 
include ISAAC on the ESO VLT UT1 \citep{isaac}, with a field of view of 
2.5\,$\times$\,2.5 arcmin$^{2}$, CISCO on the Subaru telescope,  
with a field of view 1.8\,$\times$\,2.5 arcmin$^{2}$ \citep{cisco}, and 
NIRI on the Gemini-North telescope \citep{niri}, with a 2\,$\times$\,2 
arcmin$^{2}$ field of view.  Under the same assumptions as before, an 
order of magnitude more area than \citet{tmm98} can be covered to similar 
depth in 6-10 nights.  The fields of view are comparable to the MTBW98 
survey, $\sim2\,h^{-1}$\,Mpc, and thus it may be preferable to continue 
targeting known high-redshift sources.  In the years since the MTBW98 
or \citet{tmm98} surveys were done, many new quasar damped Ly$\alpha$ 
and metal absorption line systems have been discovered, significantly 
increasing the list of potential targets. 

\acknowledgments

We would like to thank Dr. B.T. Soifer for the funding that made 
this Summer Undergraduate Research Fellowship (Caltech SURF project) 
possible.

\clearpage

\begin{deluxetable}{ccccccc}
\tabletypesize{\scriptsize}
\tablewidth{0pt}
\tablecaption{Emission-Line Objects \label{tbl-1}}
\tablehead{
   \colhead{Object}                           &
   \multicolumn{3}{c}{Magnitudes\tablenotemark{a}}                &
   \colhead{Line Flux}                       &      
   \colhead{EQW$_r$\tablenotemark{b}}         &
   \colhead{SFR\tablenotemark{c}}             \\
   \colhead{}                                 &  
   \colhead{BB$_{IR}$}                        &
   \colhead{NB$_{IR}$}                        &          
   \colhead{R$_{\rm Harris}$}                 &
   \colhead{(ergs cm$^{-2}$ s$^{-1}$)}        & 
   \colhead{(\AA)}                            & 
   \colhead{($h^{-2}$ M$_{\odot}$ yr$^{-1}$)} } 

\startdata
   Q0100$+$13B & 21.40$\pm$0.13 & 20.36$\pm$0.15 & 23.60$\pm$0.05 & $1.6\pm0.3\times 10^{-16}$ &  60$^{+21}_{-17}$ & 31$\pm$6 \\
   Q0201$+$36D & 19.59$\pm$0.08 & 18.84$\pm$0.17 & 22.87$\pm$0.03 & $1.3\pm0.2\times 10^{-16}$ &  67$^{+31}_{-24}$ & 12$\pm$2 \\
   Q2038$-$01D & 20.72$\pm$0.23 & 18.51$\pm$0.09 & 23.69$\pm$0.10 &
   $3.3\pm0.3\times 10^{-16}$ & 835$^{+591}_{-299}$ & 31$\pm$3 \\
   Q2038D-(NE) & 21.12$\pm$0.26 & 19.21$\pm$0.21 & \nodata        & $1.6\pm0.3\times 10^{-16}$ & 488$^{+100}_{-25}$  & 15$\pm$3 \\
   Q2038D-(SW) & 21.63$\pm$0.39 & 18.75$\pm$0.14 & \nodata        &
   $2.8\pm0.4\times 10^{-16}$ & $>11,000$\tablenotemark{d}                                                              & 26$\pm$4  
\enddata

\tablenotetext{a}{Q0100+13B images taken using $J$ and 1.237$\mu$m
  filters. Q0201+36D \& Q2038-01D images taken using $K$ and
  2.248$\mu$m filters.}
\tablenotetext{b}{Rest Frame Equivalent Width}
\tablenotetext{c}{Star Formation Rate}
\tablenotetext{d}{The derived equivalent width is highly nonlinear at
  colors $>$ 2 magnitudes, approaching infinity for ($K - 2.248\mu$m) $\rightarrow$ 2\fm96.}
\end{deluxetable}

\end{document}